\documentclass[preprint]{aastex}
\usepackage{color}
\pdfoutput=1

\title{A Search for Hydroxylamine (NH$_2$OH) toward Select Astronomical Sources}
\author{Robin L. Pulliam}
\affil{National Radio Astronomy Observatory, Charlottesville, VA 22903, USA}
\author{Brett A. McGuire}
\affil{Division of Chemistry and Chemical Engineering, California Institute of Technology\\ Pasadena, CA 91125, USA}
\author{Anthony J. Remijan}
\affil{National Radio Astronomy Observatory, Charlottesville, VA 22903, USA}

\begin{document}

\begin{abstract}
Observations of 14 rotational transitions of hydroxylamine (NH$_2$OH) using the NRAO 12 m Telescope on Kitt Peak are reported towards IRC+10216, Orion KL, Orion S, Sgr B2(N), Sgr B2(OH), W3IRS5, and W51M.  Although recent models suggest the presence of NH$_2$OH in high abundance, these observations resulted in non-detection.  Upper limits are calculated to be as much as six orders of magnitude lower than predicted by models.  Possible explanations for the lower than expected abundance are explored. \\
\end{abstract}

\section{Introduction}

Hydroxylamine (NH$_2$OH) has been suggested as a possible reactant precursor in the formation of interstellar amino acids \citep{Blagojevic2003,Snow2007}.   The presence of amino acids in the gas-phase toward astronomical environments would have a profound impact on the effort to understand the origin of complex molecular material in space.  The recent discovery of the simplest amino acid glycine (NH$_2$CH$_2$COOH) in cometary samples collected by the Stardust mission has provided new clues towards our knowledge of the delivery of prebiotic material to planetesimals \citep{Elsila2009} yet not to their initial formation.  That is, are complex prebiotic molecules, such as glycine, formed via reactions of smaller precursors \textit{after} their incorporation into cometary bodies, or do these complex molecules form first in the gas phase before accretion?  The search for interstellar, gas-phase glycine has therefore attracted much attention, but has yet to be unambiguously confirmed in space \citep{Kuan2003,Snyder2005}.  Recently, laboratory experiments have shown that ionized NH$_2$OH, reacting in the gas phase with acetic acid (CH$_3$COOH) and propanoic acid (CH$_3$CH$_2$COOH), can lead to the formation of glycine and the amino acids $\alpha$- and $\beta$-alanine (CH$_3$CH(NH$_2$)COOH) \citep{Blagojevic2003,Snow2007}. As acetic acid has already been observed in various environments (Shiao et al.\ 2010 and references therein), the detection of NH$_2$OH would be of much interest to the astrochemistry community, helping to answer the question of how these large complex molecules form in astronomical environments.

Protonated hydroxlyamine, NH$_3$OH$^+$ and NH$_2$OH$_2$$^+$, is also fundamentally interesting as a prebiotic molecule, having been shown to be a precursor to amino acid formation \citep{Snow2007}.  As shown in Equations \ref{Eq1}-\ref{Eq3}, protonated hydroxylamine can react with CH$_3$COOH (Equation \ref{Eq1}) and CH$_3$CH$_2$COOH (Equations \ref{Eq2} and \ref{Eq3}) to produce protonated glycine and protonated $\beta$- and $\alpha$-alanine, respectively. Since CH$_3$COOH is a well established interstellar molecule, the detection of NH$_3$OH$^+$ in the ISM would greatly enhance our understanding of the possible formation route to glycine and possibly other simple amino acids in interstellar environments \citep{Mehringer1997}. 
\begin{equation}
\mbox{NH}_3\mbox{OH}^+ + \mbox{CH}_3\mbox{COOH} \rightarrow \mbox{NH}_3\mbox{CH}_2\mbox{COOH}^+ + \mbox{H}_2\mbox{O}
\label{Eq1}
\end{equation}
\begin{equation}
\mbox{NH}_3\mbox{OH}^+ + \mbox{CH}_3\mbox{CH}_2\mbox{COOH} \rightarrow
\mbox{NH}_3\mbox{CH}_2\mbox{CH}_2\mbox{COOH}^+ + \mbox{H}_2\mbox{O}
\label{Eq2}
\end{equation}
\begin{equation}
\mbox{NH}_3\mbox{OH}^+ + \mbox{CH}_3\mbox{CH}_2\mbox{COOH} \rightarrow
\mbox{CH}_3\mbox{CH(NH}_3\mbox{)COOH}^+ + \mbox{H}_2\mbox{O}
\label{Eq3}
\end{equation}

There are few laboratory studies of NH$_2$OH or protonated NH$_2$OH formation.  Nishi et al.\ (1984) proposed a route for synthesis of NH$_2$OH involving ice mixtures of water and ammonia where a radical recombination reaction (Equation \ref{Eq4}) just above the surface of the ice under irradiated conditions produces NH$_2$OH.
\begin{equation}
\mbox{NH}_2 + \mbox{H} + \mbox{H}_2\mbox{O} \rightarrow \mbox{NH}_2\mbox{OH} + \mbox{H}_2
\label{Eq4}
\end{equation}

Additionally, Zheng \& Keiser (2010) have recently produced NH$_2$OH through electron irradiation of water-ammonia ices.  They propose that NH$_2$OH results from the radical recombination of NH$_2$ and OH inside the ices.  The results of both of these studies suggest that radical reactions within ice mantles on grain surfaces may be responsible for NH$_2$OH production.

In fact, two recent gas-grain chemical models employ such reactions of radicals in their simulations.  Charnley et al.\ (2001) assumed that nitrogen atoms will first react with OH in the gas-phase to produce large amounts of NO (Equation \ref{Eq5}).  A fraction of NO ($\sim$10\%) is then accreted onto dust grains where it can then be converted to form species such as HNO and NH$_2$OH through H addition reactions.  This formation pathway is contingent upon the depletion of NO onto dust grains in significant quantities in astronomical environments, though solid evidence for NO on grain surfaces is limited.  Observational data for the presence of NO on grain mantles was first reported in the infrared by Allamandola \& Norman (1978) via the fundamental rovibrational band at 5.3 $\mu$m.  More recently, Akyilmaz et al.\ (2007) have shown that gas-phase NO is depleted towards the peak of dust emission in two sources, suggesting that NO has accreted onto the grains in these regions.
\begin{equation}
\mbox{N} + \mbox{OH} \rightarrow \mbox{NO} + \mbox{H}
\label{Eq5}
\end{equation}

Modeling by Garrod et al.\ (2008) employs a more expansive network of radical-radical reactions within the ice-mantle, incorporating large radicals formed from photolysis of the ice constituents already known to be present. These radical ``fragments" go on to react, building more complex species as they become mobile on the grain surface through a gradual warm-up process before being liberated into the gas phase. Formation of NH$_2$OH is predicted to start from the radical-radical reaction of NH + OH addition on grain surfaces followed by hydrogenation or directly by the reaction of OH + NH$_2$.  The model predicts an NH$_2$OH column density as high as 10$^{16}$ cm$^{-2}$; easily within the detectable limits of modern radio telescopes.

Given the potential importance of NH$_2$OH to prebiotic chemistry and the high predicted abundances, we conducted a search for NH$_2$OH in the frequency range of 130-170 GHz towards seven sources: IRC+10216, Orion KL, Orion S, Sgr B2(N), Sgr B2(OH), W3IRS5, and W51M.  While these sources are known to contain copious amounts of complex molecular material, no definitive evidence was found for NH$_2$OH toward any of these sources.  Upper limits to the beam averaged column density of NH$_2$OH were calculated based on the 1$\sigma$ rms noise limit of the observed spectra and we discuss possible explanations for the lower than expected abundances.

\section{Observations}

A 2 mm spectral line survey of IRC+10216, Orion KL, Orion S, Sgr B2(N), Sgr B2(OH), W3IRS5, and W51M (hereafter, the Turner 2mm Survey) was conducted using the NRAO\footnote{The National Radio Astronomy Observatory (NRAO) is a facility of the National Science Foundation, operated under cooperative agreement by Associated Universities, Inc.} 12 m telescope on Kitt Peak by B. E. Turner between 1993 and 1995\footnote{The survey data are available online (http://www.cv.nrao.edu/Turner2mmLineSurvey) with the Spectral Line Search Engine (SLiSE) developed by A. J. Remijan and M. J. Remijan.  Further details of the Turner 2mm survey including the motivation for a complete survey of these sources are described in Remijan et al.\ 2008, arXiv:0802.2273v1 [astro-ph]}.  Table 1 lists the observing parameters for each source in the survey.  Column 1 lists the source; columns 2 and 3 list the reported pointing positions precessed to J2000 coordinates\footnote{The precession from B1950 to J2000 coordinates was done using the FUSE Precession Routine available at: $\rm{http://fuse.pha.jhu.edu/support/tools/precess.html}$}; column 4 reports the observed spectral linewidth in km s$^{-1}$ and column 5 lists the assumed source local standard of rest velocity (km s$^{-1}$) for each source.  The frequency range covered by this survey was between 130-170 GHz and the half-power beam width (HPBW) varied from 38\arcsec -- 46\arcsec across the band. The observations were taken using a dual channel, SIS junction single side band receiver with typical receiver noise ranging from 75 - 100 K.  The backend consisted of a 768 channel, 600 MHz bandwidth hybrid spectrometer with spectral resolution of 0.781 MHz per channel or $\sim$1.3 km/s at 150 GHz.  The intensity scale at the NRAO 12m is given as $T_R^*$ and corrects for forward spillover loss.  The radiation temperature is defined in Equation \ref{Eq6}, where $\zeta _c$ is the beam efficiency. These data were mined for the all the available 2 mm lines of NH$_2$OH listed in Table 2.

\begin{equation}
T_R=T_R^*/\zeta _c
\label{Eq6}
\end{equation}

In total, 54 transitions of NH$_2$OH are reported between 130 and 170 GHz from the published literature \citep{Muller2005,Morino2000,Tsunekawa1972}.  Of these 54 transitions, 14 were selected, five \textit{a}-type transitions and nine \textit{c}-type transitions, in this search for having the largest line strength and lowest upper state energy level.  Table 2 is a summary of each searched transition and lists the NH$_2$OH transition rest frequency (column 1), the transition quantum numbers (column 2), $\theta_b$ is the telescope beam size at the observed frequency (column 3), $E_u$ is the upper state energy (column 4), the transition type (column 5), Log$_{10}$($A_{ij}$) is the Einstein A coefficient (column 6) and g$_{J_u}$ is the $J$ degeneracy of the upper state (column 7).  Other relevant spectroscopic parameters such as the NH$_2$OH dipole moments, partition function and rotational constants are listed in the Notes of Table 2.

\section{Results}

\begin{figure}
\caption{Observed \textit{a}-type transitions of NH$_2$OH are simulated in red over the observed spectrum in black.  Simulated spectra are shown divided by a factor of 100.  An unscaled simulation is shown in blue for Sgr B2(N) for illustrative purposes.}
\plotone{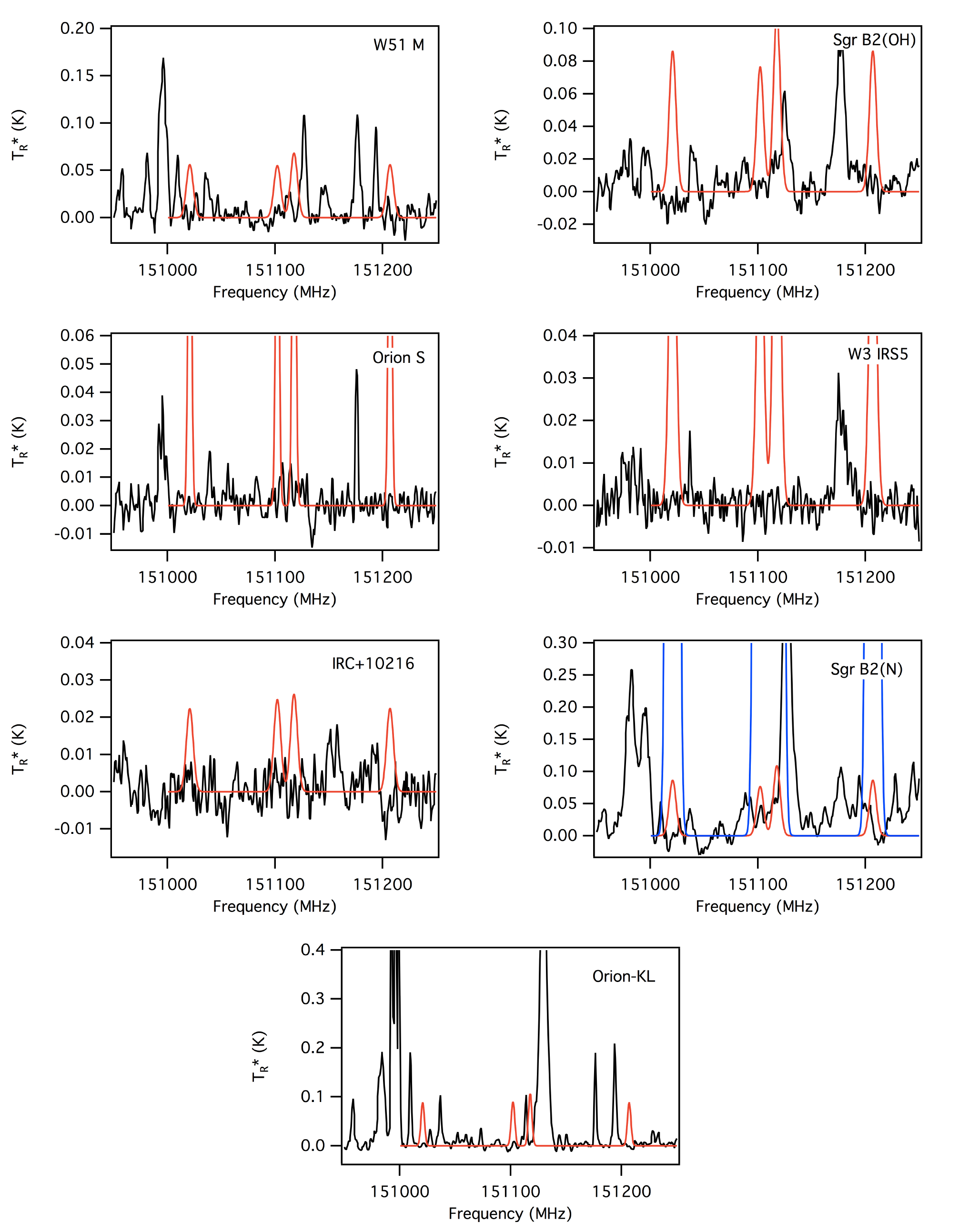}
\label{atype}
\end{figure}

\begin{figure}
\caption{Observed \textit{c}-type transitions of NH$_2$OH are simulated in red over the observed spectrum in black.   No scaling factor has been applied to the simulated spectra.}
\plotone{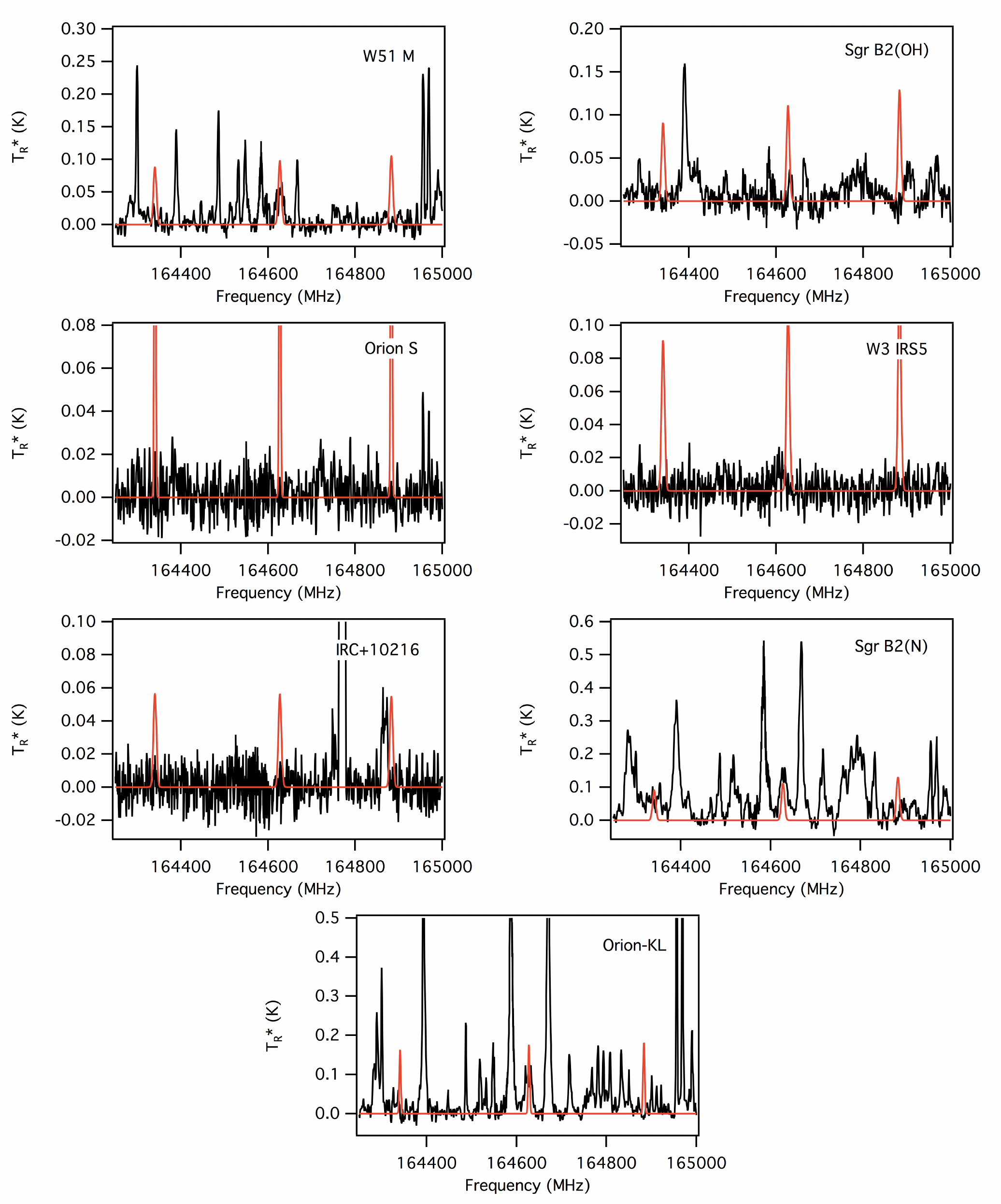}
\label{ctype1}
\end{figure}

\begin{figure}
\caption{Observed \textit{c}-type transitions of NH$_2$OH are simulated in red over the observed spectrum in black.  No scaling factor has been applied to the simulated spectra.}
\plotone{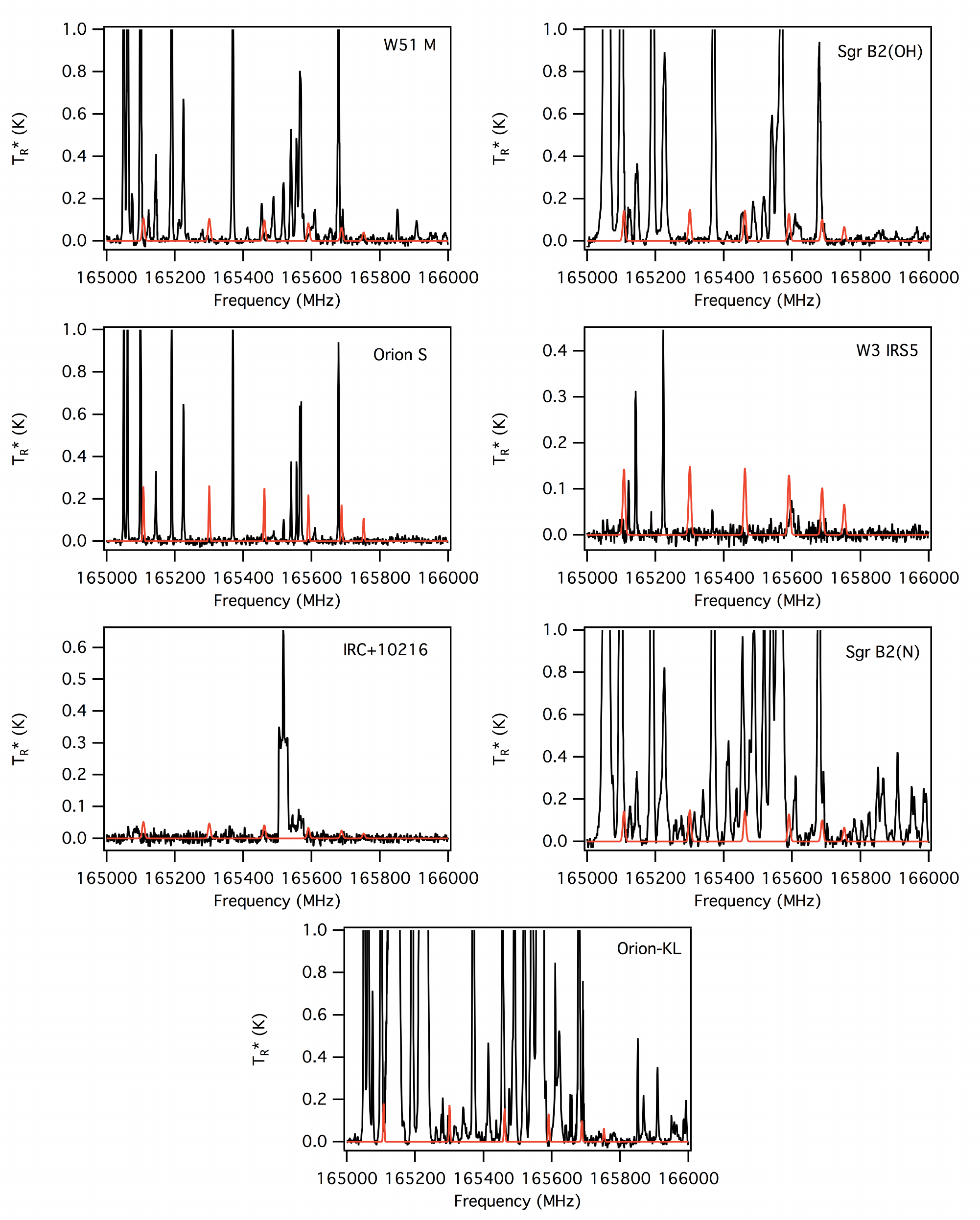}
\label{ctype2}
\end{figure}

Figures \ref{atype}-\ref{ctype2} show the observed spectra (black trace) for each source in the frequency range of the NH$_2$OH target transitions\footnote{For the purposes of this study, we have not identified the other molecular features within Figures 1-3. The data are publically available (refer to Footnote 2) and we encourage the readers to download the data and use the publically available spectral line databases for assignments.}.  Shown in red is a simulated spectrum of the expected transition line strengths from NH$_2$OH using the total column density predicted for Sgr B2(N) from the Garrod et al.\ (2008) model and source-appropriate rotational temperatures and line widths (Equation 4).  The simulated spectra in red has been scaled down by a factor of 100 for clarity. For illustrative purposes, an unscaled simulation is shown in blue for Sgr B2(N) in Figure 1. While the total column density of NH$_2$OH in sources other than Sgr B2(N) can be expected to vary based on chemical composition and physical environment, the simulations serve to show in a qualitative sense that the searched for transitions of NH$_2$OH are not present toward these sources beyond the 1$\sigma$ RMS noise limit and certainly not present in the abundances predicted by the model. In several sources, emission features are present at a number of the appropriate center frequencies for NH$_2$OH.  Yet, the strongest transitions of NH$_2$OH in this band are the $J$ = 3-2 manifold near 151 GHz and in no source are all of the expected transitions observed.  This indicates that the observed emission features are coincidental overlap with other molecular transitions and NH$_2$OH is not observed in these sources. 

Using Equation \ref{Eq7}, upper limits to the beam averaged column density based on the 1$\sigma$ RMS noise limit were calculated and are reported in Table \ref{cds}.  An approximate rotational temperature appropriate for each source was used based on data available in the references shown in Table \ref{cds}. For the purposes of this work, CH$_3$OH was used as a primary source of temperature information when available.  CH$_3$OH was chosen as it is a well known temperature probe and traces the molecular gas of the desired regions. Additionally, the wealth of observational data on CH$_3$OH establishes confidence in temperatures derived from its observations.  It should be noted that the upper limits presented here are fairly insensitive to the relatively minor range of rotational temperatures observed in these sources. Fractional abundances with respect to molecular hydrogen were calculated for each source based on these upper limits.  In the case of Sgr B2(N), the Garrod et al.\ (2008) model predicts a relative abundance for NH$_2$OH of 3.5 x 10$^{-7}$ - 4.2 x 10$^{-6}$, which is up to six orders of magnitude higher than the observed upper limits (see Table 3). 

\begin{equation}
N_{tot}=\frac{8\pi \nu ^2 k Q_{rot} \int T_R^* dV}{g_{J_u}g_{k_u}hc^3A}e^{E_u/kT_{rot}}
\label{Eq7}
\end{equation}

\section{Discussion}

In this paper we reported on the negative detection of hydroxylamine (NH$_2$OH) towards several astronomical sources.  Upper limits to the beam averaged column density have also been determined for each source based on the 1$\sigma$ RMS noise level in each spectra.  Recent chemical models introduced a new gas-grain chemical network utilizing radical-radical reactions as formation mechanisms (Garrod et al.\ 2008).  The model reproduces the beam averaged column densities of species such as methanol (CH$_3$OH), acetaldehyde (CH$_3$CHO), and even glycolaldehyde (CH$_2$(OH)CHO) with excellent agreement with current observed abundances toward the Sgr B2(N) star-forming region \citep{Garrod2008}.  However, for NH$_2$OH, the predicted abundances are 3.5$\times$10$^{-7}$ - 4.2$\times$10$^{-6}$, nearly six orders of magnitude higher than the observed upper limit of 8$\times$10$^{-12}$ towards Sgr B2(N) reported in this study.  The following sections discuss possible explanations for this surprising difference, focusing on possible formation and destruction mechanisms.

\subsection{Formation Mechanisms}

The formation of NH$_2$OH within ice grains was first proposed by Nishi et al.\ (1984) as shown in Equation \ref{Eq1}.  Previous experimental attempts to produce NH$_2$OH within the gas phase through the reaction of HNO$^+$ with H$_2$ have failed \citep{Blagojevic2003,Lias1988}.  The Garrod et al.\ (2008) grain chemistry model assumes two formation mechanisms for NH$_2$OH, both of which assume radical-radical reactions within the grain mantle.  In early times, NH$_2$OH is formed through the barrierless (see Figure 1 of Garrod et al.\ 2008) reactions of the hydroxyl radical (-OH) with NH followed by hydrogenation (Equations \ref{Eq8} and \ref{Eq9}) similar to the well studied hydrogenation reactions of CO forming CH$_3$OH \citep{Woon2002}. 
\begin{equation}
\mbox{NH} + \mbox{OH} \rightarrow \mbox{HNOH}
\label{Eq8}
\end{equation}
\begin{equation}
\mbox{HNOH} + \mbox{H} \rightarrow \mbox{NH}_2\mbox{OH}
\label{Eq9}
\end{equation}
However, there is no current theoretical or experimental work to suggest this hydrogenation reactionleading to the final product of hydroxlyamine in equation (9) proceeds in a manner similar to CO as the model assumes.  In fact, given the different states of these molecules, CO having a $^1\Sigma ^+$ electronic configuration while NO is a $^2\Pi$, it is likely that the hydrogenation of NO will proceed quite differently from that of CO.  This difference could account for the higher abundances predicted by the Garrod et al.\ (2008) model at lower temperatures. It is also worth bearing in mind that the branching ratios found in the gas phase are not applicable to solid state processes as the surrounding ice can carry away the energy fairly efficiently.

As warming takes place and the hydroxyl radical becomes more mobile on the surface of the grain, the model predicts the barrierless reaction with NH$_2$ to become dominant (Equation \ref{Eq10}). 
\begin{equation}
\mbox{NH}_2 + \mbox{OH} \rightarrow \mbox{NH}_2\mbox{OH}
\label{Eq10}
\end{equation}
However, experimental studies have shown \citep{Schnepp1960} that in an isolated argon matrix, NH$_2$ quickly combines with a free hydrogen radical to form NH$_3$ when a temperature of 20 K was reached. The question then becomes whether NH$_2$ and OH have a higher probability to react to form NH$_2$OH in interstellar ices before they recombine with free hydrogen to form NH$_3$ and H$_2$O, respectively.

Interstellar ices are considerably more complex than the isolated matrices used in the Schnepp \& Dressler (1960) laboratory study, as are the ices considered in the Garrod et al.\ (2008) model which contain a number of other simple species (e.g. CH$_4$, CH$_3$OH, NH$_3$, CO, CO$_2$, HCOOH, H$_2$O).  As such, NH$_2$ and OH are not the only species present to react with free hydrogen which may instead react with itself (to form H$_2$) or with other smaller species.   An examination of the rates of reaction of free hydrogen with these species might therefore be in order to help determine if NH$_2$ and OH would be available in sufficient quantities to collide and form enough NH$_2$OH to be detectable in the ISM. However, the diffusion and evaporation rates of atomic H and other radicals are of more importance. Radical-radical reactions will proceed at the diffusion rates of the reactants and evaporation dominates the loss channels of atomic hydrogen under almost all temperature regimes, which therefore makes it the most important route for determining overall hydrogen populations.    Unfortunately, these diffusion rates of reactants are difficult to ascertain experimentally. 

There is experimental evidence to support the radical-radical formation of NH$_2$OH from NH$_2$ and OH precursors in electron-irradiated ammonia-water ice samples \citep{Zheng2010}. Upon irradiation, an NH$_3$ species is found to undergo unimolecular decomposition to form the NH$_2$ radical and a free hydrogen atom (Equation 11). Water decomposes in a similar fashion, forming OH and H.
\begin{equation}
\mbox{NH}_3 \rightarrow \mbox{NH}_2 + \mbox{H}
\label{Eq11}
\end{equation}

  After irradiation, a new absorption peak at $\sim$1500 cm$^{-1}$ was observed and attributed to NH$_2$ formation.  As the ice samples were warmed, the species released in the gas phase were monitored by IR and mass spectroscopy.  The presence of NH$_2$OH was first noted in the IR measurements as the sample reached 174 K.  As the temperature continued to rise, the abundance of NH$_2$OH decreased until non-detection at 200 K.  NH$_2$OH was also observed in the mass spectroscopy measurements from 160 - 180 K.  It is important to note that NH$_2$OH is observed at temperatures above which most, if not all, of the water and ammonia are sublimed. 

While this study does support the formation of NH$_2$OH through radical-radical recombination within interstellar ices, it also provides potential evidence as to why NH$_2$OH is not currently observed in the ISM.  Is there a possible temperature problem?  The Garrod et al.\ (2008) model predicts the high abundance of NH$_2$OH with a temperature on the order of $\sim$130 K.  

Zheng \& Keiser (2010) find that water, ammonia and hydroxlyamine co-desorb, while some NH$_2$OH remains in the ice samples to somewhat higher temperatures. According to the experimental data, NH$_2$OH was not observed in the gas phase until temperatures exceeded 160 K. However, temperature programmed laboratory desorption experiments demonstrate behavior that is strongly dependent on the heating rate and this experimentally determined evaporation temperature does not correspond to the actual interstellar value. Without running futher analyses of the laboratory data, such as fitting the data to a Polanyl-Wigner type expression (see Galway \& Brown 1999 and references therein) or running chemical models, making a qualitative comparison between evaportation temperatures from the lab and in the interstellar medium is difficult.

Also, strong water absorption bands obscure NH$_2$OH absorption features, making its detection using infrared observations towards hot core regions impossible. However, it is possible the NH$_2$OH emission is confined to very compact ($<$5$''$) hot core regions such as the SgrB2(N-LMH) where the temperatures are higher and hydroxlyamine would be completely released from the grain surfaces \citep{Miao1995}. Single dish observations from this survey would be too beam diluted to detect the emission from this compact region. Such an occurence has been noted before.  Acetic acid (CH$_3$COOH) has been confirmed in several hot core regions using interfermetric observations after searches with single dish observations resulted in negative detections \citep{Wooten1992, Mehringer1997, Remijan2002, Remijan2003}. Additionally, many of the single dish spectra presented here are very dense with molecular emission. Interferometric images have less line confusion than that of single dish. Many of features in single dish spectra arise from an extended region or from different locations within the source and this is demonstrated with the detection of amino acetonitrile by Belloche et al. (2008).  The single dish spectra of amino acetonitrile were heavily contaminated with other complex molecular species with only 88 of the 398 observed transitions being reported as relatively "clean". Interferometric observations of Sgr B2(N) confirmed the compact emission of amino acetonitrile with a source size of 2" \citep{Belloche2008}.  As such, higher spatial resolution interferometric observations may be needed to more thoroughly couple to the higher temperature regions in order to detect hydroxlyamine.

Alternatively, observations could be conducted towards molecular sources in shocked regions such as the bipolar outflow L1157(B). In these types of sources, molecules which are formed on grain surfaces but which are not liberated into the gas phase by thermal desorption due to low temperatures are instead ejected into the gas phase by shocks \citep{Requena2006}.  Detection of NH$_2$OH in these sources would provide valuable insight into the mechanisms behind its formation pathways and eventual release into the gas phase.

\subsection{Protonated Hydroxylamine}

Next we examine possible pathways for the destruction of NH$_2$OH once it enters the gas phase. It is well known that ion-molecule reactions are important in gas phase interstellar chemistry and that protonated species play an important role in reaction mechanisms.  NH$_2$OH, having a high proton affinity ($\sim$193.5 kcal mol$^{-1}$) is particularly susceptible to protonation from other species such as H$_3^+$, HCO$^+$, CH$_5^+$, H$_3$O$^+$ and CH$_3$OH$_2^+$ (Blagojevic et al.\ 2003 and references therein).  The energies of protonation of NH$_2$OH by H$^+$ and the possibility of proton transfer by H$_3^+$ have been predicted by theory \citep{Boulet1999,Angelelli1995,Perez1998}.  As a result of protonation, two stable species were reported: NH$_3$OH$^+$ and NH$_2$OH$_2^+$, with NH$_3$OH$^+$ found to be more stable by $\sim$100 kJ mol$^{-1}$.

The reaction of NH$_2$OH with either H$^+$ or protonated methanol (CH$_3$OH$_2^+$)  was predicted to be very exothermic and it was proposed that the excess energy would either dissociate the species or could lead to the rearrangement of the species to the higher energy NH$_2$OH$_2^+$.  This could result in an enhanced abundance of NH$_2$OH$_2^+$ in the ISM.   Once in the gas phase, recent theoretical work has shown that the reaction of ionized and protonated NH$_2$OH with H$_2$, its most likley collision partner, is highly unfavorable \citep{Largo2009}.  These species, therefore, are likely to remain as reaction partners for further chemistry.

Given these considerations, even if NH$_2$OH is produced on ice grains through radical-radical reactions, upon its release into the gas phase it may quickly undergo protonation.  This would result in very low observed abundances of NH$_2$OH in the ISM.  Once protonated, the reaction with H$_2$, by far the most likely collision partner, is highly unfavorable and the lifetimes of these species should therefore be greatly enhanced.  A search for NH$_2$OH$_2^+$ and NH$_3$OH$^+$ within these star forming regions might be prudent; although, dissociative recombination reactions could result in lowered abundances of these species.  This would first require the acquisition of the rotational spectra of these species in the laboratory to enable astronomical searches.

\section{Conclusion}

We report the non-detection of NH$_2$OH towards seven sources.  Calculated upper limits for the abundance of this molecule are as much as six orders of magnitude lower than those predicted for the species by recent models.  Several factors could account for this discrepancy including the rapid removal of precursor molecules from ice mantles through reaction with free hydrogen or the rapid protonation (and subsequent dissociative recombination) of NH$_2$OH by H$^+$, H$_3^+$, CH$_5^+$, and other efficient protonation mechanisms.  The single dish observations presented here are likely highly beam diluted.  Higher resolution interferometric observations could provide the sensitivity required for detection and therefore allow better refinement of models which currently predict the presence of NH$_2$OH in high abundance.

\acknowledgments

BAM gratefully acknowledges G.A. Blake for his support, as well as funding by an NSF Graduate Research Fellowship. We also would like  to thank the anonymous reviewer for the valuable comments and suggestions to improve the
quality of the paper.

\begin{deluxetable}{l r r c c}
\tablewidth{0pt}
\tablecaption{Observed sources and coordinates (J2000), spectral line widths, and $v_{LSR}$ for each source.}
\tablecolumns{5}
\tablehead{\colhead{Source} & \colhead{RA} & \colhead{Dec} & \colhead{$\Delta V$} & \colhead{$v_{LSR}$} \\
\colhead{} & \colhead{} & \colhead{} & \colhead{(km s$^{-1}$)} & \colhead{(km s$^{-1}$)}}
\startdata
Orion KL   	& 05 35 14.5 	& -05 22 32.6		& 4.5		& $+9^a$ 	\\
Orion S       	& 05 35 16.5 	& -05 19 26.7 		& 4.5 	        & $+7^b$	\\
IRC+10216 	& 09 47 57.3 	&  +13 16 43.0 		& 9.0		& $-26^c$		\\
Sgr B2(OH)  	& 17 47 20.8 	& -28 23 32.2 		& 9.0		& $+60^d$	\\
Sgr B2(N)    	& 17 44 11.0 	& -28 22 17.3 		& 9.0		& $+62^e$		\\
W51M        	& 19 16 43.8 	&  +14 30 07.5 		& 9.0		& $+57^f$	\\
W3IRS5     	& 02 27 04.1 	&  +61 52 21.4 		& 9.0		& $-39^g$		\\
\enddata
\vspace{-2em}
\tablerefs{ \\
 a) \citep{Ziurys1993} \\
 b) \citep{Menten1988} 
 c) \citep{Latter1996} \\
 d) \citep{Turner1991} 
 e) \citep{Nummelin2000} \\
 f) \citep{Millar1988} 
 g) \citep{Helmich1994} \\
}
\label{sources}
\end{deluxetable}

\begin{deluxetable}{c c c c c c c}
\tablecaption{Observed transitions of NH$_2$OH, beam size, and parameters used to simulate the spectra and calculate NH$_2$OH beam averaged column densities (see Equation \ref{Eq4}).}
\tablewidth{0pt}
\tablecolumns{7}
\tablehead{\colhead{Rest Frequency} & \colhead{Transition} & \colhead{$\theta _b$} & \colhead{E$_u$} & \colhead{Type} & \colhead{Log$_{10}$($A_{ij}$)} & \colhead{g$_{J_u}$}\\
\colhead{(MHz)} & \colhead{$J_u(K_aK_c)-J_l(K_aK_c)$} & \colhead{($\prime \prime$)} & \colhead{(cm$^{-1}$)} & \colhead{} & \colhead{(s$^{-1}$)} & \colhead{}}
\startdata
151020.70&	3(1,3)-2(1,2)&  41.52	&	10.57&	a	& -5.27587	&	7	\\
151101.99&	3(2,2)-2(2,1)& 	41.50 	&	27.16&	a	& -5.47928	&	7	\\
151102.32&	3(2,1)-2(2,0)&	41.50	&	27.16&	a	& -5.47928	&	7	\\
151117.67&	3(0,3)-2(0,2)&	41.49	&	5.04&	a	& -5.22390	&	7	\\	
151207.01&	3(1,2)-2(1,1)&	41.47	&	10.57&	a	& -5.27422	&	7	\\
164340.78&	9(1,9)-9(0,9)&	38.15	&	75.59&	c	& -7.02982	&	19	\\
164627.49&	8(1,8)-8(0,8)&	38.09	&	60.48&	c	& -7.02783	&	17	\\
164883.24&	7(1,7)-7(0,7)&	38.03	&	47.04&	c	& -7.02617	&	15	\\
165107.71&	6(1,6)-6(0,6)&	37.98	&	35.28&	c	& -7.02472	&	13	\\
165300.63&	5(1,5)-5(0,5)&	37.93	&	25.2&	c	& -7.02335	&	11	\\
165461.76&	4(1,4)-4(0,4)&	37.89	&	16.8&	c	& -7.02237	&	9	\\
165590.89&	3(1,3)-3(0,3)&	37.87	&	10.08&	c	& -7.02148	&	7	\\
165687.89&	2(1,2)-2(0,2)&	37.84	&	5.04&	c	& -7.02079	&	5	\\
165752.62&	1(1,1)-1(0,1)&	37.83	&	1.68&	c	& -7.02037	&	3	\\
\enddata
\vspace{-2em}
\tablecomments{
 a) Molecular data were obtained from the Cologne Database for Molecular Spectroscopy \citep{Muller2005} available at www.splatalogue.net \citep{Remijan2008b}.  The uncertainties of the transition frequencies are 50 kHz \citep{Morino2000}.\\
 b) Degeneracies calculated as: $g_{J_u}=2J+1$, $g_{K_u=0}=1$, $g_{K_u\ne 0}=2$ \\
 c) Rotational constants from the CDMS Database: \\ A = 190976.2 MHz, B = 25218.73 MHz, C = 25156.66 MHz \\
 d) NH$_2$OH dipole moments in Debye \citep{Muller2005}: $\mu_A$=0.589; $\mu_C$=0.060 \\ 
 e) The functional form of the rotational partition function was determined from Equation 3.69 of Gordy \& Cook (Third Ed., 1984) - Q$_{rot}$=0.5$T_{rot}^{1.5}$ and confirmed by a fit to the partition function data given in \citep{Muller2005}.
}
\label{transitions}
\end{deluxetable}

\begin{deluxetable}{l c l l l c}
\tablecaption{1$\sigma$ RMS level of $T_R^*$, upper limits on column density of NH$_2$OH, H$_2$ column density, relative abundance of NH$_2$OH, and assumed value of $T_{rot}$.}
\tablecolumns{6}
\tablewidth{0pt}
\tablehead{ 
\colhead{} & \colhead{$T_R^*$} & \colhead{$N_{NH_2OH}$} & \colhead{$N_{H_2}$} & \colhead{$N_{NH_2OH}/N_{H_2}$} & \colhead{$T_{rot}$} \\
\colhead{Source} & \colhead{(mK)} & \colhead{(cm$^{-2}$)} & \colhead{(cm$^{-2}$)} & \colhead{}&  \colhead{(K)}
}
\startdata
Orion KL$^a$    & 6.4 & $<$2 x 10$^{13}$ & 7.0 x 10$^{23}$  & $<$3 x 10$^{-11}$ & 120\\
Orion S$^b$       & 4.0 & $<$9 x 10$^{12}$ & 1.0 x 10$^{23}$  & $<$9 x 10$^{-11}$ & 80 \\
IRC+10216$^c$ & 5.1 & $<$8 x 10$^{13}$ & 3.0 x 10$^{22}$  & $<$3 x 10$^{-9}$ & 200 \\
Sgr B2(OH)$^d$  & 7.0 & $<$3 x 10$^{13}$ & 1.0 x 10$^{24}$  & $<$3 x 10$^{-11}$ & 70 \\
Sgr B2(N)$^e$    & 6.6 & $<$2 x 10$^{13}$ & 3.0 x 10$^{24}$ & $<$8 x 10$^{-12}$ & 70  \\
W51M$^f$        & 7.7 & $<$4 x 10$^{13}$ & 1.0 x 10$^{24}$  & $<$4 x 10$^{-11}$ & 100 \\
W3IRS5$^g$    & 4.3 & $<$2 x 10$^{13}$ & 5.0 x 10$^{23}$ & $<$3 x 10$^{-11}$ & 70 \\
\enddata
\vspace{-2em}
\tablerefs{ \\
 a) T$_{rot}$ from \citep{Ziurys1993}; N$_{H_2}$ from \citep{Womack1992} and Refs. therein. \\
 b) T$_{rot}$ from \citep{Menten1988}; N$_{H_2}$ from \citep{Womack1992} and Refs. therein. \\
 c) T$_{rot}$ from \citep{Patel2011}; N$_{H_2}$ from \citep{Cernicharo2010}\\
 d) T$_{rot}$ from \citep{Turner1991}; N$_{H_2}$ from \citep{Womack1992} and Refs. therein. \\
 e) T$_{rot}$ and  N$_{H_2}$ from \citep{Nummelin2000} \\
 f) T$_{rot}$ from \citep{Millar1988}; N$_{H_2}$ from \citep{Womack1992} and Refs. therein. \\
 g) T$_{rot}$ and  N$_{H_2}$ from \citep{Helmich1994} \\
}
\label{cds}
\end{deluxetable}

\clearpage



\clearpage


\begin{thebibliography}{}

\bibitem[Akyilmaz et al. 2007]{Akyilmaz2007}Akyilmaz, M., Flower, D.R. Hily-Blant, P., Pineau es For\^{e}ts, G., \& Walmsley, C.M., 2007 \textit{Astron. \& Astrophys.}, \textbf{462}, 221.

\bibitem[Allamandola \& Norman 1978]{Allamandola1978}Allamandola, L.J. \& Norman, C.A., 1978 \textit{Astron. \& Astrophys.}, \textbf{63}, L23.

\bibitem[Angelelli et al. 1995]{Angelelli1995}Angelelli, F., Aschi, M., Cacace, F., Pepi, F., \& de Petris, G., 1995 \textit{J. Phys. Chem.}, \textbf{99}, 6551.

\bibitem[Belloche et al. 2008]{Belloche2008}Belloche, A., Menten, K.M., Comito, C., M\"{u}ller, H.S.P., Schilke, P. , Ott, J., Thorwirth, S., \& Hieret, C., 2008, \textit{Astron. \& Astrophys.}, \textbf{482}, 179.

\bibitem[Blagojevic et al. 2003]{Blagojevic2003}Blagojevic, V., Petrie, S., \& Bohme, D.K., 2003 \textit{M. N. R. A. S.}, \textbf{339}, L7.

\bibitem[Boulet et al. 1999]{Boulet1999}Boulet, P., Gilardoni, F., Weber, J., Chermette, H., \& Ellinger, Y., 1999 \textit{Chem. Phys.}, \textbf{244}, 163.

\bibitem[Cernicharo et al. 2010]{Cernicharo2010}Cernicharo, J., Waters, L.B.F.M., Decin, L., \& Encrenaz, P. et al., 2010 \textit{Astron. \& Astrophys.}, \textbf{521}, L8.

\bibitem[Charnley et al. 2001]{Charnley2001}Charnley, S.B., Rodgers, S.D., \& Ehrenfreund, P., 2001 \textit{Astron. \& Astrophys.}, \textbf{378}, 1024.

\bibitem[Elsila et al. 2009]{Elsila2009}Elsila, J.E., Glavin, D.P., \& Dworkin, J.P., 2009 \textit{Meteor. \& Planet. Sci.}, \textbf{44}, 1323.

\bibitem[Flower et al. 2005]{Flower2005}Flower, D.R., Pineau des Forets, G., \& Walmsley, C.M., 2005 \textit{Astron. \& Astrophys.}, \textbf{436}, 933.

\bibitem[Galway \& Brown 1999]{Galway1999}Galway, A.K, \& Brown, B.E. 1999, Thermal Decomposition of Ionic Solids, Elsevier: New York, NY, 123. 
\bibitem[Garrod et al. 2008]{Garrod2008}Garrod, R.T., Widicus Weaver, S.L., \& Herbst, E., 2008 \textit{Astrophys. J.}, \textbf{682}, 283.

\bibitem[Gordy \& Cook (Third Ed, 1984)]{Gordy}Gordy, W., \& Cook, R.L. 1984, Microwave Molecular Spectra, Third Ed., Wiley-Interscience: New York, NY, 58.

\bibitem[Helmich et al. 1994]{Helmich1994}Helmich, F.P., Jansen, D.J., de Graauw, Th., Groesbeck, T.D., \& van Dishoeck, E.F., 1994 \textit{Astron. \& Astrophys.}, \textbf{283}, 626.

\bibitem[Herbst \& Leung 1986]{Herbst1986}Herbst, E. \& Leung, C. M., 1986 \textit{M. N. R. A. S.}, \textbf{222}, 689.

\bibitem[Kuan et al. 2003]{Kuan2003}Kuan, Y.J., Charnley, S.B., Huang, H.C., Tseng, W.L., \& Kisiel, Z., 2003 \textit{Astrophys. J.}, \textbf{593}, 848.

\bibitem[Largo et al. 2009]{Largo2009}Largo, L., Ray\'{on}, V.M., Barrientos, C., Largo, A., \& Redondo, P., 2009 \textit{Chem. Phys. Lett.}, \textbf{476}, 174.

\bibitem[Latter \& Charnley 1996]{Latter1996}Latter, W.B. \& Charnley, S.B., 1996 \textit{Astrophys. J.}, \textbf{463}, L37.

\bibitem[Lias et al. 1988]{Lias1988}Lias, S.G., Barmess, J.E., \& Liebman, J.F. et al., 1988 \textit{J. Phys. Chem. Ref. Dat.}, \textbf{17 (Supplement 1)}, 1.

\bibitem[McGonagle et al. 1990]{McGonagle1990}McGonagle, D., Ziruyz, L.M., Irvine, W. M., \& Minh, Y. C., 1990 \textit{Astrophys. J.}, \textbf{359}, 121.

\bibitem[Mehringer et al. 1997]{Mehringer1997}Mehringer, D.M., Snyder, L.E., Miao, Y.T., \& Lovas, F.J., 1997 \textit{Astrophys. J.}, \textbf{480}, L71.

\bibitem[Menten et al. 1988]{Menten1988}Menten, K.M., Walmsley, C.M., Henkel, C., \& Wilson, T.L., 1988 \textit{Astron. \& Astrophys.}, \textbf{198}, 253.

\bibitem[Miao et al. 1995]{Miao1995}Miao, Y., Mehringer, D.M., Kuan, Y., \& Snyder, L.E., 1995 \textit{Astrophys. J.}, \textbf{445}, L59.

\bibitem[Millar et al. 1988]{Millar1988}Millar, T.J., Olofsson, H., Hjalmarson, \r{A}., \& Brown, P.D., 1988, \textit{Astron. \& Astrophys.}, \textbf{205}, L5.

\bibitem[Morino et al. 2000]{Morino2000}Morino, I., Yamada, K.M.T., Klein, H., \& Belov, S.P. et al., 2000 \textit{J. Mol. Struct.}, \textbf{517-518}, 367.

\bibitem[M\"{u}ller et al. 2005]{Muller2005}M\"{u}ller, H.S.P., Schl\'{o}der, F., Stutzki, J., \& Winnewisser, G., 2005 \textit{J. Mol. Struct.}, \textbf{742}, 215.

\bibitem[Nishi et al. 1984]{Nishi1984}Nishi, N., Shinohara, H., \& Okuyama, T., 1984 \textit{J. Chem. Phys.}, \textbf{80}, 3898.

\bibitem[Nummelin et al. 2000]{Nummelin2000}Nummelin, A., Bergman, P., \& Hjalmarson, \r{A}., et al., 2000 \textit{Astrophys. J. Suppl.}, \textbf{128}, 213.

\bibitem[Patel et al. 2011]{Patel2011}Patel, N.A, Young, K.H., \& Gottlieb, C.A., et al., 2011 \textit{Astrophys. J. Suppl.}, \textbf{193}, 
17.

\bibitem[P\`{e}rez \& Contreras 1998]{Perez1998}P\`{e}rez, P. \& Contreras, R., 1998 \textit{Chem. Phys. Lett.}, \textbf{293}, 239.

\bibitem[Ragan et al. 2011]{Ragan2011}Ragan, S.E., Bergin, E.A., \& Wilner, D., 2011 \textit{Astrophys. J.}, \textbf{736}, 163.

\bibitem[Remijan et al. 2002]{Remijan2002}Remijan, A.J., Snyder, S.-Y.L., Liu, S.-Y., Mehringer, D., \& Kuan, Y.-J., 2002 \textit{Astrophys. J.}, \textbf{576}, 264. 

\bibitem[Remijan et al. 2003]{Remijan2003}Remijan, A.J., Snyder, S.-Y.L., Friedel, D.N., Liu, S.-Y., \& Shah, R.Y., 2003 \textit{Astrophys. J.}, \textbf{590}, 314. 

\bibitem[Remijan et al. 2008]{Remijan2008}Remijan, A.J., Hollis, J.M., Lovas, F.J., Stork, W.D., Jewell, P.R., \& Meier, D.S., 2008 \textit{Astrophys. J.}, \textbf{675}, L85.

\bibitem[Remijan \& Markwick-Kemper 2008]{Remijan2008b}Remijan, A.J., \& Markwick-Kemper, A.J., 2008, \textit{Bull. of the Am. Astron. Soc.}, \textbf{39}, 963.

\bibitem[Requena-Torres et al. 2006]{Requena2006}Requena-Torres, M.A., Martin-Pintado, J., Rodriguez-Franco, A., Martin, S., Rodriguez-Fernandez, N.J., \& de Vincente, P., 2006 \textit{Astron. \& Astrophys.}, \textbf{455}, 971.

\bibitem[Schnepp \& Dressler 1960]{Schnepp1960}Schnepp, O. \& Dressler, K., 1960 \textit{J. Chem. Phys.}, \textbf{32}, 1682.

\bibitem[Shiao et al. 2010]{Shiao2010}Shiao, Y.S.J., Looney, L.W., Remijan, A.J., Snyder, L.E., \& Friedel, D.N., 2010 \textit{Astrophys. J.}, \textbf{716}, 286.

\bibitem[Snow et al. 2007]{Snow2007}Snow, J.L., Orlova, G., Blagojevic, V., \& Bohme, D.K., 2007 \textit{J. Am. Chem. Soc.}, \textbf{129}, 9910.

\bibitem[Snyder et al. 2005]{Snyder2005}Snyder, L.E., Lovas, F.J., \& Hollis, J.M., et al., 2005 \textit{Astrophys. J.}, \textbf{619}, 914.

\bibitem[Tsunekawa 1972]{Tsunekawa1972}Tsunekawa, S., 1972 \textit{J. Phys. Soc. Japan}, \textbf{33}, 167.

\bibitem[Turner 1991]{Turner1991}Turner, B.E., 1991 \textit{Astrophys. J. Suppl.}, \textbf{76}, 617.

\bibitem[Woon 2002]{Woon2002}Woon, D.E., 2002 \textit{Astrophys. J.}, \textbf{569}, 541.

\bibitem[Wooten et al. 1992]{Wooten1992}Wooten, A., Wlodarczack, G., Mangum, J.G., Combes, F., Encrenaz, P.J., \& Gerin, M., 1992 \textit{Astron. \& Astrophys.}, \textbf{257}, 740.

\bibitem[Zheng \& Kaiser 2010]{Zheng2010}Zheng, W.J. \& Kaiser, R.L., 2010 \textit{J. Phys. Chem. A.}, \textbf{114}, 5251.

\bibitem[Ziurys \& McGonagle 1993]{Ziurys1993}Ziurys, L.M. \& McGonagle, D., 1993 \textit{Astrophys. J. Suppl.}, \textbf{89}, 155.

\bibitem[Womack et al. 1992]{Womack1992}Womack, M., Ziurys, L.M., \& Wyckoff, S., 1992 \textit{Astrophys. J.}, \textbf{393}, 188.

\end{thebibliography}
\end{document}